\documentclass[aps, prb, reprint]{revtex4-1}

\pdfoutput=1 
\usepackage{graphicx}
\usepackage{color,amsmath}
\usepackage[normalem]{ulem}

\usepackage[allcolors=blue,colorlinks,linkcolor=blue,citecolor=blue,urlcolor=blue]{hyperref}

\usepackage{siunitx}
\usepackage{physics}
\usepackage{comment}
\usepackage{layouts}

\begin{document}

\title{Gate-tunable Electronic Transport in p-type GaSb Quantum Wells}

\author{Matija Karalic}
\email{makarali@phys.ethz.ch}
\affiliation{Solid State Physics Laboratory, ETH Zurich, 8093 Zurich, Switzerland}

\author{Christopher Mittag}
\affiliation{Solid State Physics Laboratory, ETH Zurich, 8093 Zurich, Switzerland}

\author{Michael Hug}
\affiliation{Solid State Physics Laboratory, ETH Zurich, 8093 Zurich, Switzerland}

\author{Kenji Shibata}
\affiliation{Tohoku Institute of Technology, Sendai 982-8577, Japan}

\author{Thomas~Tschirky}
\affiliation{Solid State Physics Laboratory, ETH Zurich, 8093 Zurich, Switzerland}

\author{Werner Wegscheider}
\affiliation{Solid State Physics Laboratory, ETH Zurich, 8093 Zurich, Switzerland}

\author{R.~Winkler}
\affiliation{Department of Physics, Northern Illinois University, DeKalb, Illinois 60115, USA}

\author{Klaus Ensslin}
\affiliation{Solid State Physics Laboratory,  ETH Zurich, 8093 Zurich, Switzerland}

\author{Thomas Ihn}
\affiliation{Solid State Physics Laboratory, ETH Zurich, 8093 Zurich, Switzerland}

\date{\today}

\begin{abstract}
We investigate two-dimensional hole transport in GaSb quantum wells at cryogenic temperatures using gate-tunable devices. Measurements probing the valence band structure of GaSb unveil a significant spin splitting of the ground subband induced by spin-orbit coupling. We characterize the carrier densities, effective masses and quantum scattering times of these spin-split subbands and find that the results are in agreement with band structure calculations. Additionally, we study the weak anti-localization correction to the conductivity present around zero magnetic field and obtain information on the phase coherence. These results establish GaSb quantum wells as a platform for two-dimensional hole physics and lay the foundations for future experiments in this system. 
\end{abstract}

\maketitle

\section{Introduction}

GaSb is a III-V narrow-bandgap binary semiconductor that posseses high bulk hole mobility at room temperature, and is consequently of technological importance in electronics and optoelectronics \cite{dutta_physics_1997, bennett_antimonide-based_2005}. Low-power, high-performance p-type field-effect transistors (FETs) based on GaSb for complementary logic are under development. Efforts to realize such FETs are either based on thin surface channels of GaSb in a typical metal-oxide-semiconductor (MOS) configuration \cite{ali_fermi_2010, merckling_gasb_2011, nainani_optimization_2011}, or on modulation doped heterostructures hosting GaSb quantum wells (QWs) \cite{bennett_strained_2008, tokranov_algaassb_2011, bennett_enhanced_2013}. In the latter case, no field effect has been reported so far. Meanwhile, GaSb p-type nanowires are also the subject of intense research, much for the same reasons \cite{jeppsson_characterization_2008, burke_growth_2010, borg_synthesis_2013, yang_surfactant-assisted_2014, yang_approaching_2015, borg_high-mobility_2017}.

From the point of view of fundamental research, \mbox{GaSb} is best known for its self-assembled quantum dots (QDs) \cite{hatami_radiative_1995, ledentsov_radiative_1995} and for forming one half of the quantum spin Hall insulator InAs/GaSb \cite{kane_$z_2$_2005, kane_quantum_2005, bernevig_quantum_2006, liu_quantum_2008, suzuki_edge_2013, du_robust_2015, mueller_nonlocal_2015}, a coupled double QW system in which hole-like states originate from the valence band of GaSb \cite{altarelli_electronic_1983, yang_evidence_1997, lakrimi_minigaps_1997, cooper_resistance_1998,  de-leon_band_1999, qu_electric_2015, karalic_experimental_2016, karalic_lateral_2017}. By themselves, GaSb QWs are unexplored at cryogenic temperatures. We have therefore set out to study two-dimensional holes confined in GaSb QWs, performing first low-temperature experiments on devices that are fully gate-tunable. Finding and analyzing Shubnikov-de Haas (SdH) oscillations, we discover a significant zero-field spin splitting of the ground subband mediated by the spin-orbit coupling (SOC), a finding supported by self-consistent $k \cdot p$ band structure calculations. The first excited subband remains inaccessible in our experiment. To characterize the spin-split subbands of the ground subband, we measure effective masses $m^*$ and quantum scattering times $\tau_\mathrm{q}$ using the temperature dependence of the SdH oscillations. Finally, we investigate the phenomenon of weak anti-localization (WAL) that occurs as a correction to the conductivity at magnetic fields close to zero, and extract the phase-coherence length.

Based on our initial, encouraging results, we believe that it should be possible to realize nanostructures in p-type GaSb QWs, all the more as material quality improves further. Fabricating QDs would be particularly interesting due to the prospects of using them for quantum computation \cite{loss_quantum_1998}. Spin qubits based on hole QDs are predicted to have long coherence times due to the weak hyperfine coupling with the nuclear spin bath and to facilitate fast control due to the SOC \cite{bulaev_spin_2005, fischer_spin_2008, fischer_hybridization_2010, szumniak_spin-orbit-mediated_2012}.   

\begin{figure}[!t]
\includegraphics[width=\columnwidth]{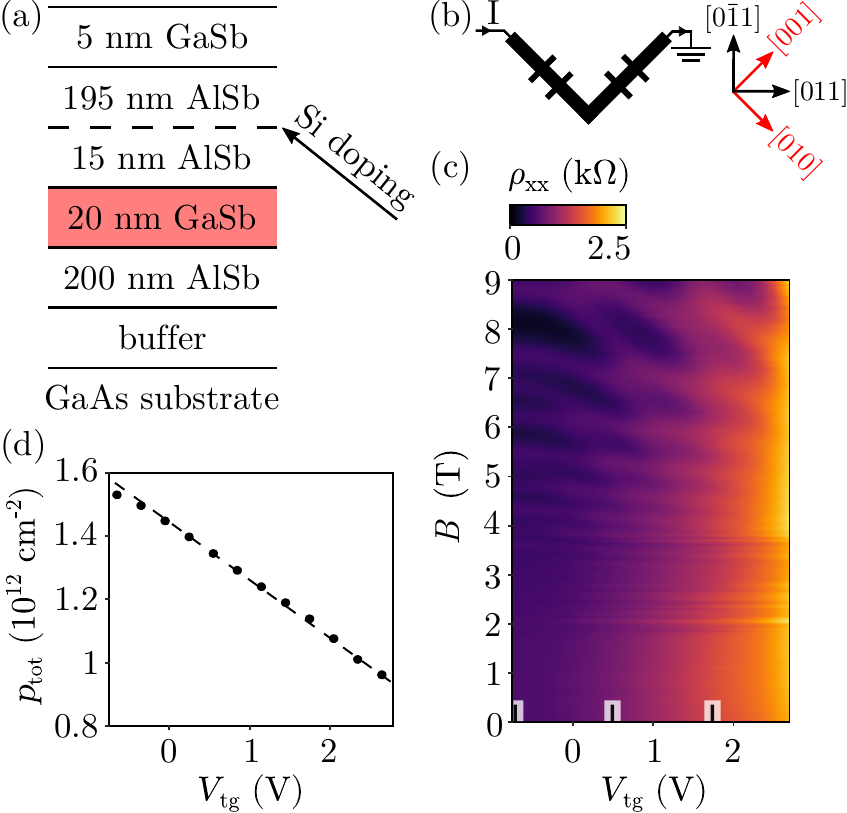}
\caption{\textbf{(a)} Composition of the heterostructure hosting the GaSb QW. The dashed line marks the position of Si dopants. \textbf{(b)} Schematic representation of the measured device and its orientation with respect to the crystallographic axes. \textbf{(c)} Longitudinal resistivity $\rho_{xx}$ as a function of top gate voltage $V_\mathrm{tg}$ and magnetic field $B$. \textbf{(d)} Hall density $p_\mathrm{tot}$ derived from the transverse resistivity $\rho_{xy}$ as a function of $V_\mathrm{tg}$ in the range $\lvert B \rvert \leq \SI{2}{\tesla}$. The dashed line is a guide to the eye indicating the linear $p_\mathrm{tot} (V_\mathrm{tg})$ dependence.}
\label{fig1}
\end{figure}

\section{Experimental Methods}

Our single-side modulation doped GaSb QWs are grown in $[100]$ direction at $\SI{540}{\celsius}$ on a GaAs substrate using molecular-beam epitaxy \cite{charpentier_molecular_2014, tschirky_mbe_2018} [see Fig.\,\ref{fig1}(a)]. An interfacial misfit buffer alleviates the resulting lattice mismatch. This buffer is composed of a GaAs layer, succeeded by AlSb and GaSb layers and an optional superlattice comprising alternating GaSb and AlSb layers. AlSb confinement barriers surround the $\SI{20}{\nano\meter}$ GaSb QW and a GaSb cap forms the surface protection layer. The Si atoms, acting as amphoteric dopants producing net p-type doping \cite{bennett_transport_2000}, are separated from the QW by a spacer. We have grown heterostructures with varying dopant sheet density $n_\mathrm{d}$ and spacer thickness $d_\mathrm{s}$, and here we present results for $n_\mathrm{d} = 18.8 \times 10^{12}$\,$\mathrm{cm}^{-2}$ and $d_\mathrm{s} = \SI{15}{\nano\meter}$.

Measurements are conducted on a gated device consisting of two Hall bar structures, each of $\SI{25}{\micro\meter}$ width and voltage probe spacing of $\SI{50}{\micro\meter}$, oriented at right angles to each other [Fig.\,\ref{fig1}(b)]. The Hall bar structures are aligned at an angle of $\SI{45}{\degree}$ to the principal crystallographic axes, and therefore parallel to the $[010]$ and $[001]$ directions, respectively. We fabricate the Hall bar structures by wet chemical etching and subsequent encapsulation by atomic layer deposition of around $\SI{25}{nm}$ of Al$_2$O$_3$, which also serves as the gate insulator for the Ti/Au top gate. Ohmic contact is achieved by indium soldering, with typical contact resistances on the order of a few tens of k$\Omega$. Measurements are performed in a dilution refrigerator at a base temperature of $\SI{80}{\milli\kelvin}$ with low-frequency lock-in techniques and constant ac current bias. We collect all data at base temperature, unless specified otherwise. All experimentally accessible observables show no dependence on the crystallographic orientation, indicating vanishing anisotropy between the $[010]$ and $[001]$ directions. We therefore display results obtained from one pair of longitudinal and one pair of transverse voltage probes only.

\section{Results}

\begin{figure}[!t]
\includegraphics[width=\columnwidth]{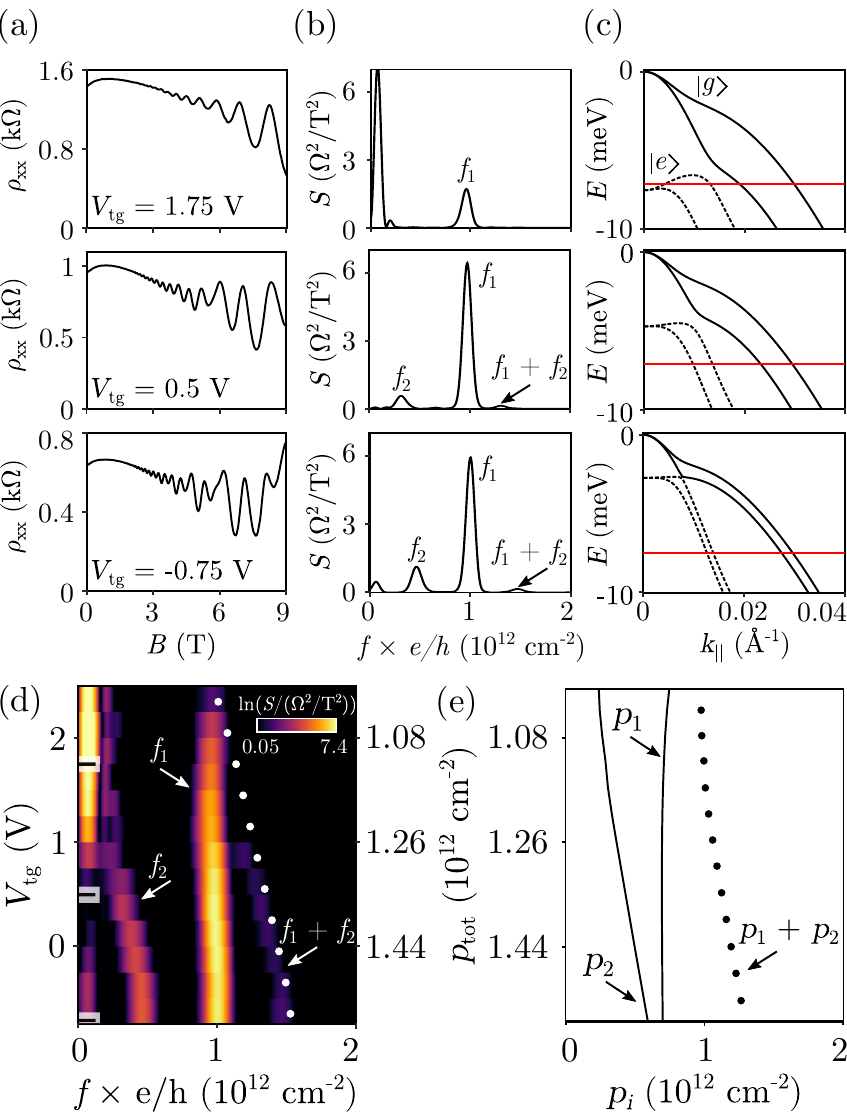}
\caption{\textbf{(a)} Examples of SdH oscillations in $\rho_{xx}$ at several values of $V_\mathrm{tg}$, as marked in Fig.\,\ref{fig1}(c). \textbf{(b)}  Power spectra $S$ obtained by Fourier transforming the traces from (a). Up to three frequencies $f_1$, $f_2$ and $f_1+f_2$ are visible, as indicated. The presence of frequencies around zero is an artifact linked to incomplete background subtraction. \textbf{(c)} Self-consistent $k \cdot p$ band structure calculations $E(k_\parallel)$ associated with the traces from (a). The total density is chosen to be the same as in (a) for each respective trace. The spin-split ground subband (solid) is labeled with $\ket g$, the spin-split first excited subband (dashed) with $\ket e$. The horizonal lines mark the position of the Fermi energy. \textbf{(d)} Collection of power spectra as in (b) in the whole $V_\mathrm{tg}$ range, presented as a color map. The dotted line follows the Hall density $p_\mathrm{tot}$ from Fig.\,\ref{fig1}(d). \textbf{(e)} Calculated evolution of the densities $p_i$ of the spin-split ground subbands with total density. Note that in the calculation a small fraction of the total density resides in the first excited subband so that $p_1 + p_2 < p_\mathrm{tot}$.}
\label{fig2}
\end{figure}

The color map in Fig.\,\ref{fig1}(c) shows the dependence of the longitudinal resistivity $\rho_{xx}$ on the top gate voltage $V_\mathrm{tg}$, which tunes the total charge carrier density, and on the perpendicular magnetic field $B$. Note the appearance of a checkerboard pattern in $\rho_{xx} (V_\mathrm{tg}, B)$ for $B > \SI{4}{\tesla}$. The Hall density $p_\mathrm{tot}$, determined from the transverse resistivity $\rho_{xy}$, represents the total charge carrier density and is depicted in Fig.\,\ref{fig1}(d). The associated Hall mobility changes from 3000 to $\SI{7500}{\centi\meter\squared\per\volt\per\second}$ in the full top gate voltage range. This is insufficient to reach the quantum Hall state with the magnetic fields at our disposal. For $V_\mathrm{tg} > \SI{2}{\volt}$, leakage through the top gate increases quickly, preventing us from attaining depletion. However, in other samples we are able to achieve full depletion, observing insulating behavior due to localization below $p_\mathrm{tot} = 5 \times 10^{11}$\,$\mathrm{cm}^{-2}$. This accomplishment, together with our measurements on nominally undoped GaSb QWs which are insulating by default due to the absence of charge carriers \cite{note1}, suggests that the edge conduction reported in InAs/GaSb double QWs \cite{nichele_edge_2016, nguyen_decoupling_2016} originates from the InAs QW, as speculated \cite{noguchi_intrinsic_1991, olsson_charge_1996, mittag_passivation_2017, de_vries_$h/e$_2018}. 

Next, we investigate $\rho_{xx}$ as function of $B$ at fixed $V_\mathrm{tg}$. Figure\,\ref{fig2}(a) shows such traces, exhibiting SdH oscillations. The corresponding power spectra $S$ calculated using the Fourier transform are displayed in Fig.\,\ref{fig2}(b). The frequency axis $f$ is converted to a density axis by multiplication with $e/h$ without making any assumptions regarding spin degeneracy. Figure\,\ref{fig2}(d) combines these and more traces into a color map, revealing the continuous evolution of the spectrum upon changing $V_\mathrm{tg}$. The dotted superimposed line describes the Hall density $p_\mathrm{tot}$, taken from Fig.\,\ref{fig1}(d). At $V_\mathrm{tg} = \SI{-0.75}{\volt}$, three distinct frequencies $f_1$, $f_2$ and $f_1 + f_2$ are visible in the spectrum. We see that $(f_1 + f_2) \times e/h$ matches $p_\mathrm{tot}$. This holds true upon increasing $V_\mathrm{tg}$ as $f_1 + f_2$ gradually disappears while moving downwards in frequency, merging with $f_1$. Accordingly, $f_2$ features a similar progression, eventually disappearing amidst the low frequencies around zero which arise from incomplete background subtraction prior to taking the Fourier transform. By $V_\mathrm{tg} = \SI{1.75}{\volt}$, $f_2$ and $f_1+f_2$ have completely vanished. $f_1$ remains almost constant throughout, decreasing only slightly with increasing $V_\mathrm{tg}$.

Based on the above observations, we attribute $f_1$ and $f_2$ to two spin-split subbands of densities $p_1 = f_1 \times e/h$ and $p_2 = f_2 \times e/h$, respectively. These subbands originate from the spin splitting of the ground subband due to the SOC. The subband-population imbalance $(p_1-p_2)/(p_1+p_2)$ measures the degree of spin splitting,  exceeding $60\%$ just before $f_2$ becomes impossible to resolve. Note that normally the densities $p_1$ and $p_2$ in a pair of spin subbands vanish jointly as a function of decreasing total density $p_1 + p_2$ due to Kramers theorem which guarantees the degeneracy of the spin subbands at $k_{\parallel} = 0$. However, it is possible that only a single spin subband is occupied below some critical density if the spin-split dispersion is characterized by a ring of extrema \cite{bychkov_1984, jo_2017}.

In order to confirm and expand upon our interpretation, we have performed self-consistent $k \cdot p$ band structure calculations based on the $8 \times 8$ Kane Hamiltonian in the axial approximation \cite{winkler_spin-orbit_2003} and using band-structure parameters from Ref.~\onlinecite{vurgaftman_2001}. The precise electrostatic boundary conditions are not known for our system.  Overall best agreement with the experimental data is achieved if we choose boundary conditions that correspond to a symmetric well for $p_\mathrm{tot} = 1.72 \times 10^{12}$\,$\mathrm{cm}^{-2}$.  Then we keep the slope of the Hartree potential in one barrier fixed, corresponding to the situation in the substrate, while we vary the slope $F$ of the Hartree potential in the other barrier, corresponding to the effect of the top gate.  The charge density $\rho(z)$ is then calculated self-consistently \cite{winkler_1993, winkler_1993a} as a function of $F$, where $z$ denotes the growth direction and $p_\mathrm{tot} = \int \rho(z) \, dz$.  Figure\,\ref{fig2}(c) shows calculated $E(k_{\parallel})$ dispersion curves for approximately the same total densities as found in the corresponding measured traces and spectra of Figs.\,\ref{fig2}(a), (b); and Fig.\,\ref{fig2}(e) shows the calculated spin subband densities $p_1$ and $p_2$ versus $p_\mathrm{tot}$.  The calculations confirm the important trends in Fig.\,\ref{fig2}(c) including the rather distinct dependencies of $p_1$ and $p_2$ on $p_\mathrm{tot}$ as well as a tremendous Rashba-type spin splitting of the ground and excited subbands. Due to the self-consistent interplay between quantum mechanical confinement and electrostatics, the band structure alters dramatically as the total density changes. The agreement with the experimental results from Fig.\,\ref{fig2}(d) attests to the validity of our conclusions. While the calculations also predict the occupation of the spin-split excited subbands for $p_\mathrm{tot} \gtrsim 1.0 \times 10^{12}$\,$\mathrm{cm}^{-2}$, we are unable to resolve this occurrence experimentally as the densities in the excited subbands are small. Note that at $k_{\parallel} = 0$ both the ground subband as well as the first excited subband have heavy hole character, whereas for typical $k_{\parallel}$ which are of the order of the Fermi wave vector, both subbands are approximately equal parts heavy hole and light hole.  

\begin{figure}[!t]
\includegraphics[width=\columnwidth]{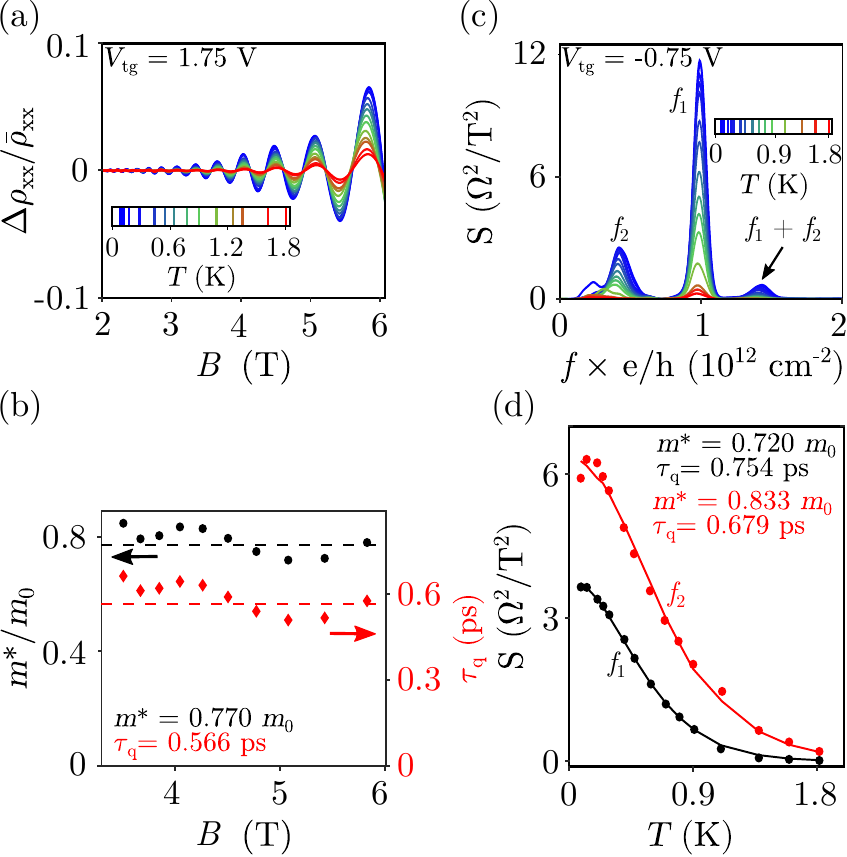}
\caption{\textbf{(a)} Temperature dependence of the relative SdH oscillation amplitude $\Delta \rho_{xx}/\bar\rho_{xx}$ at $V_\mathrm{tg} = \SI{1.75}{\volt}$ after subtracting the slowly varying background $\bar\rho_{xx}$ ($\Delta \rho_{xx} = \rho_{xx} - \bar\rho_{xx}$). \textbf{(b)} Effective mass $m^*$ and quantum scattering time $\tau_\mathrm{q}$ found from (a) with Eq.\,(\ref{eq1}) for many minima and maxima, up to $B = \SI{6}{\tesla}$. The dashed lines mark the weighted averages of $m^*$, $\tau_\mathrm{q}$, see inset. \textbf{(c)} Decay of the power spectrum at $V_\mathrm{tg} = \SI{-0.75}{\volt}$ with temperature $T$. The magnetic field range used for the Fourier transform is $\SI{2}{\tesla} \leq B \leq \SI{9}{\tesla}$ and the frequencies of interest are annotated as before. \textbf{(d)} Peak heights at frequencies $f_1$, $f_2$ as function of $T$ together with the associated fits (Method II of main text). The extracted $m^*$, $\tau_\mathrm{q}$ are inserted. The magnetic field ranges used for the Fourier transform are $\SI{2}{\tesla} \leq B \leq \SI{6}{\tesla}$ and $\SI{3}{\tesla} \leq B \leq \SI{9}{\tesla}$ in the case of $f_1$ and $f_2$, respectively.}
\label{fig3}
\end{figure}

We now turn to the temperature dependence of the \mbox{SdH} oscillations. Such measurements allow for the determination of $m^*$ and $\tau_\mathrm{q}$. At $V_\mathrm{tg} = \SI{1.75}{\volt}$ [Fig.\,\ref{fig3}(a)], only the spin-split subband of higher density, $p_1$, appears in the SdH oscillations, recall Fig.\,\ref{fig2}. We employ the well-known expression describing the oscillation amplitude $\Delta \rho_{xx}$,
\begin{equation}
\frac{\Delta \rho_{xx}}{\bar\rho_{xx}} = 4 \exp{\left(-\frac{\pi}{\omega_\mathrm{c} \tau_\mathrm{q}}\right)} \frac{2 \pi^2 k_\mathrm{B} T/\hbar \omega_\mathrm{c}}{\sinh{2 \pi^2 k_\mathrm{B} T/\hbar \omega_\mathrm{c}}},
\label{eq1}
\end{equation}
where $\bar\rho_{xx}$ is the magnetoresistance background, $\omega_\mathrm{c} = eB/m^*$ the cyclotron frequency and $T$ the temperature \cite{ando_electronic_1982}. Equation\,(\ref{eq1}) is valid for $\Delta \rho_{xx}/\bar\rho_{xx} \ll 1$. Fitting the decay of the oscillation amplitude with temperature in different minima and maxima using Eq.\,(\ref{eq1}) produces Fig.\,\ref{fig3}(b). We see that $m^*$, $\tau_\mathrm{q}$ are independent of $B$ and obtain $m^*/m_0 = 0.770 \pm 0.004$, $\tau_\mathrm{q} = (0.566 \pm 0.003)$\,ps  by weighted averaging.

Next, we repeat the temperature dependence at $V_\mathrm{tg} = \SI{-0.75}{\volt}$. Here, more intricate analysis is required because both spin-split subbands contribute to the SdH oscillations. We have applied two methods to analyze the data \cite{nichele_spin-orbit_2014}. Method I consists of filtering the data for the purpose of selecting specific frequencies, allowing for the individual analysis of the constituent components of the SdH oscillations. We first use a Butterworth bandpass filter to select $f_1$, $f_2$ or $f_1+f_2$, then replicate the analysis using Eq.\,(\ref{eq1}), as before. Disadvantages of this approach are the trade-off in filter order and frequency response as well as uncertainty in the choice of magnetic field range. Method II considers the power spectra of the SdH oscillations. We fit the heights of the \mbox{peaks} corresponding to the different subbands as function of temperature to peak heights in the power spectra of simulated data sets produced using the Fourier transform of Eq.\,(\ref{eq1}). Modifications such as choice of magnetic field range and windowing are applied to both real and simulated data sets, ensuring consistency. The fitting parameters are again $m^*$ and $\tau_\mathrm{q}$. Figure.\,\ref{fig3}(c) depicts the decay of the power spectrum with increasing temperature at $V_\mathrm{tg} = \SI{-0.75}{\volt}$. In Fig.\,\ref{fig3}(d), we show the peak heights at frequencies $f_1$ and $f_2$, corresponding to the spin-split subbands, versus temperature, together with the respective fits. For the subband of higher density, $p_1$, we obtain $m^*/m_0 = 0.720 \pm 0.006$, $\tau_\mathrm{q} = (0.754 \pm 0.007)$\,ps, and for the subband of lower density, $p_2$, $m^*/m_0 = 0.833 \pm 0.018$, $\tau_\mathrm{q} = (0.679 \pm 0.016)$\,ps. The peak associated with $f_1+f_2$ has a fictitious mass roughly equal to the sum of the other two masses, as anticipated. Method I (not shown) gives similar results as Method \mbox{II}. Comparing with the results at $V_\mathrm{tg} = \SI{1.75}{\volt}$, we conclude that $m^*$ of the higher density subband does not depend much on gate voltage, as expected from the band structure calculations. The calculations predict somewhat smaller zero-field density-of-states effective masses at the Fermi energy for both subbands than what we measure [$m^\ast / m_0 = 0.51$ for both spin subbands of the first panel in Fig.\,\ref{fig2}(c) and $m^\ast / m_0 = 0.47$ and $0.51$ for the third panel]. Such differences are known from studies in other systems and originate in the complicated structure of the valence band \cite{nichele_spin-orbit_2014}.
 
\begin{figure}[!t]
\includegraphics[width=\columnwidth]{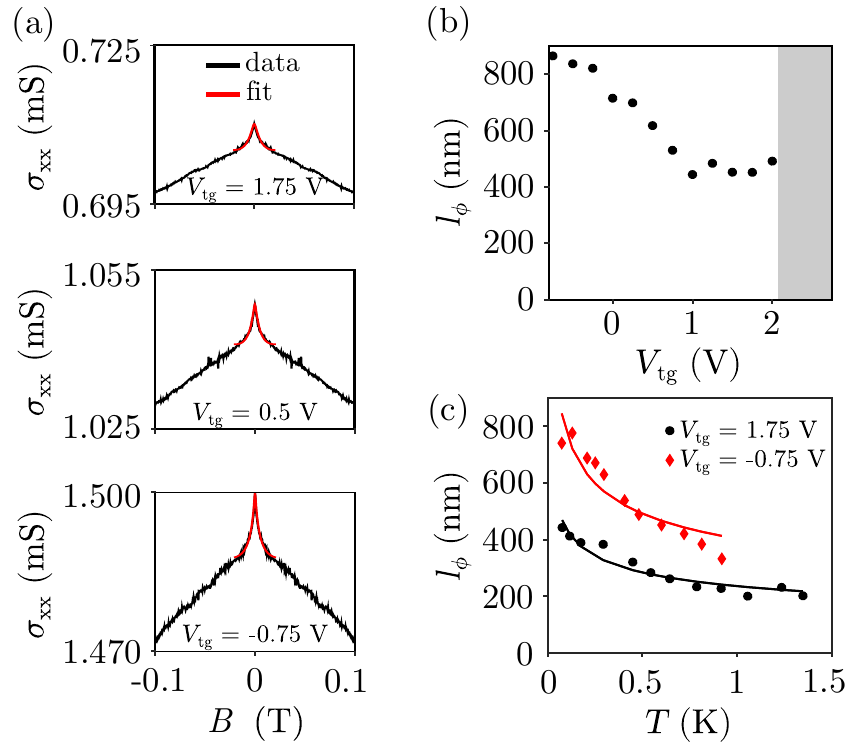}
\caption{\textbf{(a)} Examples of the WAL correction to the longitudinal conductivity $\sigma_{xx}$ at several values of $V_\mathrm{tg}$. The measured traces are shown as is, i.e., prior to any background subtraction, together with their respective fits according to Eq.\,(\ref{eq2}) in the range $\lvert B \rvert \leq \SI{20}{\milli\tesla}$. \textbf{(b)} Phase coherence length $l_\phi$ found from the WAL correction as function of $V_\mathrm{tg}$ at base temperature. In the shaded region, no reliable determination of $l_\phi$ is possible. \textbf{(c)} Temperature dependence of $l_\phi$ at $V_\mathrm{tg} = \SI{1.75}{\volt}$ and $\SI{-0.75}{\volt}$. Also depicted are fits with a power law dependence $T^\gamma$.}
\label{fig4}
\end{figure}

Our final experimental finding concerns WAL. To investigate the phenomenon of WAL, we measure the magnetic field dependence of $\rho_{xx}$ and $\rho_{xy}$ for fixed $V_\mathrm{tg}$ and calculate the longitudinal conductivity $\sigma_{xx}$ by inverting $\rho_{xx}$, $\rho_{xy}$. Then, we symmetrize and fit $\sigma_{xx}$ with a two-band model of the form $\sigma_{xx} = \sigma_{xx, 1} + \sigma_{xx, 2}$, where $\sigma_{xx, 1}$ ($\sigma_{xx, 2}$) is the conductivity of the band of density $p_1$ ($p_2$). We neglect intersubband scattering and fix the densities to $p_1 = f_1 \times e/h$ and $p_2 = f_2 \times e/h$, respectively, resulting in two fitting parameters, $\mu_1$ and $\mu_2$, the mobilities of the subbands. The two-band model gives the parabolic background explaining the observed positive magnetoresistance around $B = 0$, see for example Fig.\,\ref{fig2}(a). Note that $\rho_{xy}$ is essentially linear in $B$ despite the presence of two hole species. After subtracting the parabolic background from $\sigma_{xx}$ and getting $\delta \sigma_{xx}$, we shift the traces such that $\delta \sigma_{xx} (0) = 0$, and fit with
\begin{equation}
\resizebox{\hsize}{!}{$
\delta \sigma_{xx} = \frac{e^2}{\pi h}\left(g\left(\frac{B_\phi + B_\mathrm{so}}{B}\right) + \frac{1}{2}g\left(\frac{B_\phi + 2 B_\mathrm{so}}{B}\right) - \frac{1}{2}g\left(\frac{B_\phi}{B}\right)\right),$}
\label{eq2}
\end{equation}
where $g (x) = \Psi(1/2 + x) - \ln{x}$ and $\Psi$ is the digamma function  \cite{iordanskii_weak_1994}. The quantities $B_\phi = \hbar/4 e l_\phi^2$ and $B_\mathrm{so} = \hbar/4 e l_\mathrm{so}^2$ are the fitting parameters. We are interested in $l_\phi$, the phase coherence length, and $l_\mathrm{so}$, the spin-orbit length. Figure.\,\ref{fig4}(a) showcases several typical fits at different $V_\mathrm{tg}$. The fitting range is $\lvert B \rvert \leq \SI{20}{\milli\tesla}$. As expected, we find that $l_\phi$ is mostly independent of the fitting range, whereas $l_\mathrm{so}$ changes drastically for different ranges. This occurs because $l_\phi$ describes the well-defined peak shape of $\delta \sigma_{xx}$ around $B = 0$ where the background is flat. In contrast,  $l_\mathrm{so}$ describes the behavior of $\delta \sigma_{xx}$ away from the peak and is therefore influenced by the details of the crude background subtraction which neglects particle-particle interactions, for instance. Thus, there is a systematic error on $l_\mathrm{so}$ \cite{grbic_strong_2008, nichele_transport_2014}.

Figure.\,\ref{fig4}(b) presents the extracted $l_\phi$ for all $V_\mathrm{tg}$. $l_\phi$ increases with increasing density, reaching around $\SI{800}{\nano\meter}$ at most. We neglect to plot $l_\mathrm{so}$ in light of the discussion outlined above, but mention here that $l_\mathrm{so}$ does not seem to depend on $V_\mathrm{tg}$, and is approximately equal to $\SI{200}{\nano\meter}$.

Increasing the temperature quenches the WAL correction. Figure\,\ref{fig4}(c) illustrates the decrease of $l_\phi$ at two values of $V_\mathrm{tg}$ upon increasing the temperature. We fit the data points with a function of the form $AT^\gamma$ and obtain $\gamma = -0.27 \pm 0.02$ at $V_\mathrm{tg} = \SI{1.75}{\volt}$ and $\gamma = -0.29 \pm 0.04$ at $V_\mathrm{tg} = \SI{-0.75}{\volt}$. If electron-electron scattering were the dominant mechanism responsible for the loss of phase coherence, we would expect $\gamma = 0.5$ in a two-dimensional system such as ours \cite{chakravarty_weak_1986, ihn_semiconductor_2009}. Electron-phonon scattering is also not compatible with the noted temperature dependence. Currently, we cannot explain the unexpected value of the exponent $\gamma$.

\section{Conclusion}

To conclude, we have established GaSb QWs as a viable platform for two-dimensional hole physics. Transport measurements on gate-tunable devices, backed up by band structure calculations, reveal how the ground subband undergoes spin splitting. We determine the effective masses and quantum scattering times of these spin-split subbands using the temperature dependence of SdH oscillations. Additionally, we study the WAL phenomenon and deduce the phase coherence length. These \mbox{insights}, together with the fact that full depletion is possible, pave the way towards more elaborate experiments such as the confinement of valence band holes in QDs in GaSb. Such QDs offer the tantalizing possibility of realizing hole spin qubits that have long coherence times due to the weak hyperfine interaction and provide fast control that is enabled by the SOC. We think that with further improvements in material quality GaSb could become a viable candidate to host such spin qubits, joining the ranks of more established hole systems such as GaAs \cite{grbic_single-hole_2005, klochan_fabrication_2010, komijani_counting_2013, wang_anisotropic_2016, bogan_consequences_2017}, silicon \cite{spruijtenburg_single-hole_2013, li_single_2013, li_pauli_2015, maurand_cmos_2016, liles_spin_2018} and germanium \cite{hendrickx_gate-controlled_2018, watzinger_germanium_2018}. 


\begin{acknowledgments}
The authors acknowledge the support of the ETH FIRST laboratory and the financial support of the Swiss National Science Foundation (Schweizerischer Nationalfonds zur Förderung der Wissenschaftlichen Forschung, NCCR QSIT) and thank Mansour Shayegan for valuable discussions.
\end{acknowledgments}

%

\end{document}